\documentclass[10pt,prb,aps,superscriptaddress,twocolumn,longbibliography]{revtex4-1}
\usepackage{amssymb,amsmath}
\usepackage{graphicx}
\usepackage{color}
\usepackage{natbib}
\usepackage{hyperref}
\usepackage[utf8]{inputenc}

\newcommand{\eps}{\varepsilon}      
\newcommand{\om}{\omega}      
\newcommand{\divv}{\text{div}}	
\renewcommand{\cite}{\onlinecite}

\begin{document}

\renewcommand{\cite}[1]{[\onlinecite{#1}]}
\newcommand{\commentMaxim}[1]{{\color{red}{\it ~Maxim:~}\tt #1}}	

\title{Boundary conditions for the effective-medium description \\ of subwavelength multilayered structures}

\author{Maxim~A.~Gorlach}
\email{m.gorlach@metalab.ifmo.ru}
\affiliation{ITMO University, Saint Petersburg 197101, Russia}

\author{Mikhail Lapine}
\affiliation{School of Mathematical and Physical Sciences, University of Technology Sydney, NSW 2007, Australia}

\begin{abstract}
Nanostructures with one-dimensional periodicity, such as multilayered structures, are currently in the focus of active research in the context of hyperbolic metamaterials and photonic topological structures. An efficient way to describe the materials with subwavelength periodicity is based on the concept of effective material parameters, which can be rigorously derived incorporating both local and nonlocal responses. However, to provide any predictions relevant for applications, effective material parameters have to be supplemented by appropriate boundary conditions. In this work, we provide a comprehensive treatment of spatially dispersive bulk properties of multilayered metamaterials as well as derive boundary conditions for the averaged fields. We demonstrate that local bianisotropic model does not capture all the features related to second-order nonlocal effects in the bulk of metamaterial. As we prove, while the bulk response of multilayers does not depend on the unit cell choice, effective boundary conditions are strongly sensitive to the sequence of layers and multilayer termination. The developed theory  provides a clear interpretation of the recent experiments on the reflectance of all-dielectric deeply subwavelength multilayers suggesting further avenues to experimentally probe electromagnetic nonlocality in metamaterials.
\end{abstract}

\maketitle

\section{Introduction}

Electromagnetic properties of multilayered structures and their effective material parameters have attracted research interest since early days~\cite{Rytov}. Current fabrication techniques based on sputtering or atomic layer deposition allow one to fabricate multilayers with deeply subwavelength thickness and low roughness truly approaching the regime of metamaterial~\cite{Zhukovsky,Sukharn}. Nevertheless, a series of studies has warned against application of the standard frequency-dependent (i.e. {\it local}) effective material parameters to the multilayered metamaterials even in the subwavelength regime~\cite{Vinogradov,Elser,Orlov,Chebykin-2011}. In particular, multilayers can give rise to tri-refringence phenomenon~\cite{Orlov}, while the standard techniques of effective parameters retrieval can yield unphysical results for multilayers~\cite{Orlov-cr}.

Such peculiar behavior has recently been attributed to the {\it nonlocal (or spatially dispersive)} electromagnetic response of multilayers, which manifests itself through the dependence of polarization on electric field in the neighboring regions of space. Such electromagnetic nonlocality is described via the dependence of effective permittivity tensor $\hat{\eps}(\om,{\bf k})$ on wave vector ${\bf k}$ which is considered as a variable independent of $\om$ in order to capture linear response of the structure to the arbitrary excitation~\cite{Agranovich}. To evaluate effective nonlocal permittivity tensor, it has been proposed to use current-driven homogenization approach based on the analysis of the structure response to the external distributed currents~\cite{Silv,Alu}. The effective susceptibility is then found as a matrix which relates polarization averaged over the unit cell to the averaged electric field. Recently, this strategy has been applied to the multilayered metamaterial composed of isotropic layers of two types, and an explicit though cumbersome expression for the effective permittivity tensor has been derived for this particular case~\cite{Chebykin-2011,Chebykin-12}.

However, it is not just bulk nonlocality that determines the properties of multilayers, since boundary effects should also be taken into account~\cite{Markel-13,Lei,Maurel}. Therefore, to make the nonlocal description self-consistent, one has to supplement bulk effective permittivity by  appropriate boundary conditions. Since homogenized description includes {\it averaged} fields, boundary conditions should also be formulated for the averaged fields, and therefore it is not obvious {\it a priori} that the continuity of tangential components of electric and magnetic fields will still hold.

Furthermore, previous studies contain a clear indication that the effective boundary conditions should be modified. For instance, it has recently been suggested theoretically~\cite{Sheinfux} and verified experimentally~\cite{Zhukovsky} that the reflectance of all-dielectric multilayered structure with the layers of subwavelength thickness deviates from the predictions of the local effective medium model also depending on the structure termination. Shortly afterwards, quite a few proposals have been put forward on how to interpret this peculiar feature~\cite{Popov,Lei,Maurel,Castaldi}. Most importantly, this feature can not be explained by bulk nonlocality alone, and therefore one has to work out the correct form of boundary conditions in order to capture this phenomenon.

The rest of the article is organized as follows. In Section~\ref{sec:Bulk}, we provide a general analysis of the bulk nonlocal response of multilayered structure with arbitrary number of layers in the unit cell and derive an explicit expression for the nonlocal permittivity tensor $\hat{\eps}(\om,{\bf k})$ incorporating spatial dispersion corrections up to the second order. Using this explicit expression and the link between local and nonlocal descriptions~\cite{Skidanov,Silv}, we analyze in Sec.~\ref{sec:Permeability}, whether it is possible to formally describe all second-order spatial dispersion effects in the bulk of multilayered structure in terms of local permittivity and permeability tensors. The answer appears to be negative. In Section~\ref{sec:Boundary} we derive the effective boundary conditions for multilayered structure incorporating spatial dispersion corrections up to the second order. As we show, the form of boundary conditions appears to be sensitive to the termination of a multilayered structure, which provides an elegant interpretation of the dependence of reflectance on the sequence of layers. To illustrate our results, in Sec.~\ref{sec:Bilayer} we perform  calculations for a specific  metamaterial with two layers in the unit cell. Finally, in Sec.~\ref{sec:Outlook} we discuss the obtained results concluding with the outlook for future applications.

\section{Derivation of the bulk nonlocal response}\label{sec:Bulk}

\begin{figure}[b]
\begin{center}
\includegraphics[width=0.95\linewidth]{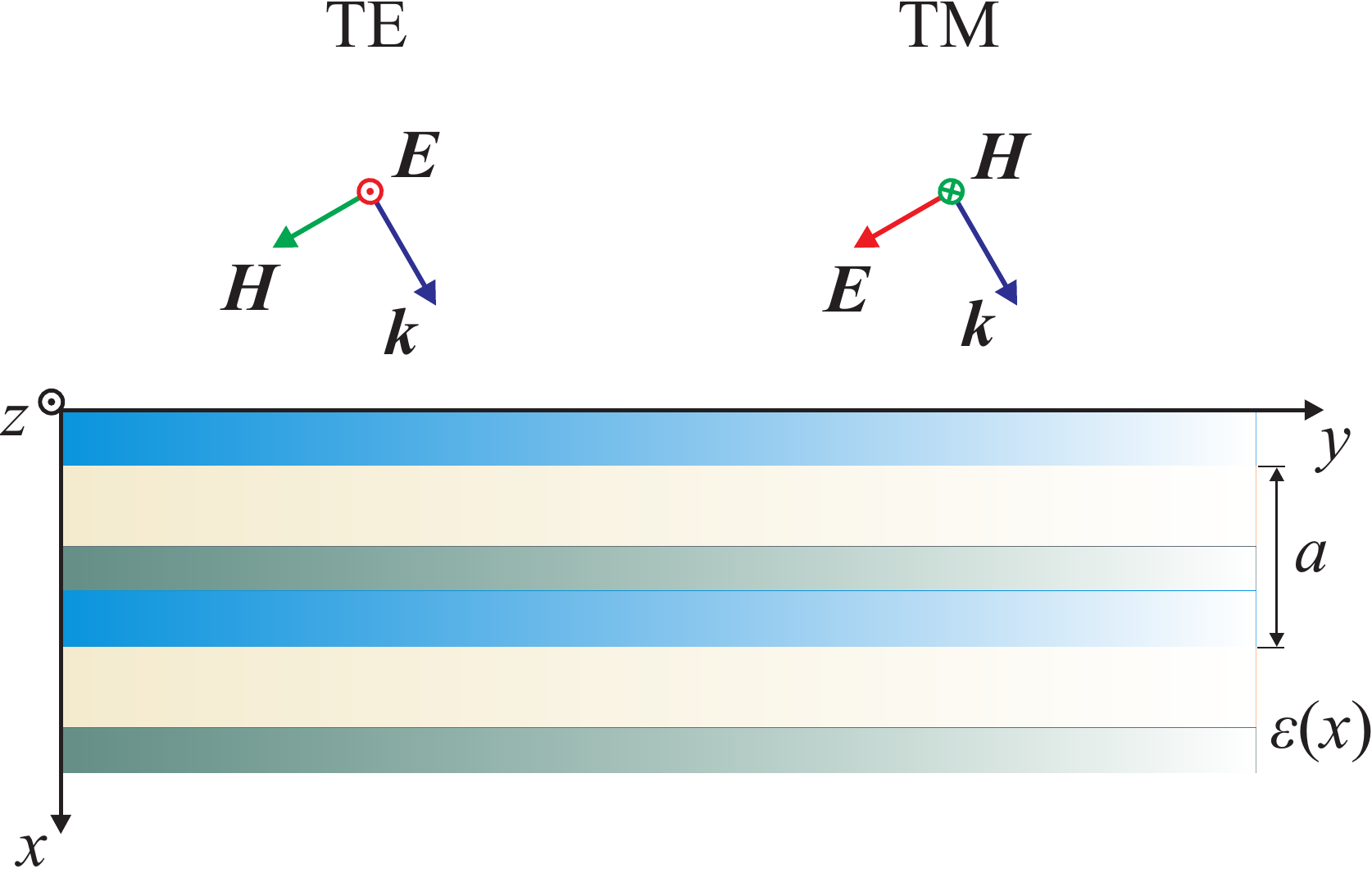}
\caption{The scheme of multilayered structure with period $a$ along $x$ axis and arbitrary permittivity profile $\eps(x)$ with two possible polarizations of plane wave sketched. For simplicity, we assume that the wave vector ${\bf k}$ defined by the external distributed currents exciting the structure lies in $Oxy$ plane.}
\label{fig:Structure}
\end{center}
\end{figure}

In this section, we provide a general derivation of the bulk response of multilayered structure assuming periodic permittivity modulation along $x$ axis, $\eps(x)$, with the period equal to $a$ as illustrated in Fig.~\ref{fig:Structure}. Analogously to Refs.~\cite{Chebykin-2011,Chebykin-12}, we use nonlocal homogenization approach assuming excitation of the structure by the external distributed currents 
\begin{equation}\label{CurrentDistr}
{\bf j}({\bf r})={\bf j_0}\,e^{ik_x\,x+ik_y\,y}
\end{equation}
oscillating with the frequency $\omega$. For simplicity, we assume that $\om\,a/c\ll 1$ and $ka\ll 1$, keeping the terms up to the second order in $\om\,a/c$ or $ka$. To obtain the effective permittivity expanded in terms of small parameter, we adopt the technique similar to Refs.~\cite{Rizza-2015,Rizza-2017} and expand the fields into Floquet harmonics as follows:
\begin{gather}
{\bf E}({\bf r})=e^{ik_y\,y}\,\left[{\bf E}_0\,e^{ik_x\,x}+\sum\limits_{n\not=0,n=-\infty}^{\infty}{\bf E}_n\,e^{i(k_x+nb)\,x}\right]\:,\label{Efield}
\end{gather}

\begin{equation}
{\bf D}({\bf r})=e^{ik_y\,y}\,\left[{\bf D}_0\,e^{ik_x\,x}+\sum\limits_{n\not=0,n=-\infty}^{\infty}{\bf D}_n\,e^{i(k_x+nb)\,x}\right]\:,\label{Dfield}
\end{equation}
where $b=2\pi/a$ is the period of reciprocal lattice, and the dependence of the fields on $x$ coordinate is determined by the Bloch theorem. The amplitudes of Fourier coefficients ${\bf E}_n$ and ${\bf D}_n$ should be chosen in order to satisfy inhomogeneous wave equation with external current:
\begin{equation}\label{WaveEq}
\nabla(\divv{\bf E})-\Delta\,{\bf E}=q^2\,{\bf D}+{\bf J}\:,
\end{equation}
where $q=\omega/c$, ${\bf J}({\bf r})\equiv 4\pi\,i\,q\,{\bf j}({\bf r})/c$ and CGS system of units is employed.

Besides that, using the structure periodicity, we expand permittivity $\eps(x)$ and inverse permittivity $\zeta(x)\equiv 1/\eps(x)$ into the Fourier series:
\begin{gather}
\eps(x)=\eps_0+\sum\limits_{n\not=0,n=-\infty}^{\infty}\,\eps_n\,e^{inbx}\:,\label{PermittivityExpansion}\\
\zeta(x)=\zeta_0+\sum\limits_{n\not=0,n=-\infty}^{\infty}\,\zeta_n\,e^{inbx}\:,\label{InvPermittivityExpansion}
\end{gather}
where the coefficients $\eps_n$ and $\zeta_n$ are related to each other due to the fact that $\eps(x)\,\zeta(x)\equiv 1$. We aim to calculate the effective permittivity tensor of a multilayered structure defined as
\begin{equation}
{\bf D}_0=\hat{\eps}(\om,{\bf k})\,{\bf E}_0\:.
\end{equation}
Note that the amplitudes ${\bf E}_0$ and ${\bf D}_0$ precisely correspond to the averaged fields used in metamaterial homogenization procedure~\cite{Silv}, while the obtained $\hat{\eps}(\om,{\bf k})$ corresponds to effective permittivity definition in any periodic medium~\cite{Agranovich}. It is also assumed that the external current ${\bf J}$ has only zeroth Floquet harmonic.

Combining the expansions Eqs.~\eqref{Efield}, \eqref{Dfield} with Eq.~\eqref{WaveEq}, we get a linear system of equations for ${\bf E}_n$ and ${\bf D}_n$ coefficients. The entire system splits into two independent sets of equations which correspond to TE and TM-polarized waves propagating in the structure. We analyze TE case below, whereas more involved case of TM polarization is examined in Appendix A.

In case of TE polarization, wave equation~\eqref{WaveEq} and the material equation ${\bf D}({\bf r})=\eps(x)\,{\bf E}({\bf r})$ yield:
\begin{gather}
\left(k_x^2+k_y^2\right)\,E_{0z}=q^2\,D_{0z}+J_z\:,\label{FourierTE1}\\
\left[(k_x+nb)^2+k_y^2\right]\,E_{nz}=q^2\,D_{nz}\mspace{10mu}(n\not=0)\:,\label{FourierTE2}\\
D_{0z}=\eps_0\,E_{0z}+\sum\limits_{n\not=0}\,\eps_{-n}\,E_{nz}\:,\label{FourierTE3}\\
D_{nz}=\eps_n\,E_{0z}+\sum\limits_{m\not=-n}\,\eps_{-m}\,E_{(n+m)z}\:,\mspace{10mu}(n\not=0)\label{FourierTE4}
\end{gather}
where the summation is performed from $-\infty$ to $\infty$. As a first step, we examine the quasistatic case when $q=k=0$. Inspecting Eq.~\eqref{FourierTE2}, we recover that $E_{nz}=0$. Then Eq.~\eqref{FourierTE3} yields $D_{0z}=\eps_0\,E_{0z}$ and the $zz$ component of effective permittivity is simply equal to the average permittivity $\eps_0$.

Hence, in the limit $qa\ll 1$ and $ka\ll 1$ we can consider $E_{nz}$ as small parameter. To the leading order, Eq.~\eqref{FourierTE2} yields that for $n\not=0$
\begin{equation*}
E_{nz}\approx\frac{q^2}{n^2\,b^2}\,D_{nz}\:.
\end{equation*}
$D_{nz}$ which enters the right-hand side can now be evaluated as $D_{nz}\approx \eps_n\,E_{0z}$ [Eq.~\eqref{FourierTE4}], since $E_{nz}$ with nonzero $n$ are already small. Thus, the expression for $E_{nz}$ reads:
\begin{equation}\label{Enz0}
E_{nz}\approx \frac{q^2\,\eps_n}{n^2\,b^2}\,E_{0z}\:.
\end{equation}
More detailed analysis incorporating frequency and spatial dispersion effects up to the third order carried on in Appendix A yields a bit more precise result for $E_{nz}$ Floquet harmonic:
\begin{equation}\label{Enz}
E_{nz}\approx \frac{q^2\,\eps_n}{n^2\,b^2}\,E_{0z}-\frac{2k_x\,q^2\,\eps_n}{n^3\,b^3}\,E_{0z}\:.
\end{equation}
Now we make use of Eqs.~\eqref{FourierTE3} writing $D_{0z}$ in terms of $E_{0z}$ as follows:
\begin{equation}\notag
D_{0z}=\eps_0\,E_{0z}+\frac{q^2}{b^2}\,\sum\limits_{n\not=0} \frac{\eps_{-n}\,\eps_n}{n^2}\,E_{0z}-\frac{2\,k_x\,q^2}{b^3}\,\sum\limits_{n\not=0}\,\frac{\eps_{-n}\,\eps_n}{n^3}\,E_{0z}\:.
\end{equation}
Since the sum $\sum_{n\not=0}\,\eps_{-n}\,\eps_n/n^3$ vanishes due to cancellation of $n$ and $-n$ terms, the result for the effective permittivity reads
\begin{equation}\label{Epsilon-zz}
\eps_{zz}(q,{\bf k})= \eps_{||}\equiv \eps_0+\frac{q^2}{b^2}\,\sum\limits_{n\not=0}\,\frac{\eps_{-n}\,\eps_n}{n^2}\:.
\end{equation}
Equation~\eqref{Epsilon-zz} suggests that spatial dispersion effects do not affect $\eps_{zz}$ component of permittivity at least up the third order, and only frequency dispersion is manifested. This is consistent with the conclusions of Refs.~\cite{Elser,Lei,Maurel}. This also agrees with the conclusions of Ref.~\cite{Orlov}, where the authors do not observe any pronounced manifestations of nonlocality for TE polarized waves.

The analysis for TM-polarized waves appears to be more involved and provided in Appendix A. One of the key differences from TE case is the emergence of the nonzero off-diagonal components $\eps_{xy}$ and $\eps_{yx}$. Their symmetric part $(\eps_{xy}+\eps_{yx})/2$ appears to be proportional to the product $k_x\,k_y$, and it is this term which has been interpreted in Ref.~\cite{Chebykin-2011} as the rotation of the optical axis of metamaterial induced by spatial dispersion. At the same time, an antisymmetric part $(\eps_{xy}-\eps_{yx})/2$ is {\it linear} in wave vector component $k_y$, being nonzero only in the case when mirror symmetry of the unit cell is broken for any unit cell choice. This effect predicted in Ref.~\cite{Rizza-2015} is not possible for bi-layer structures since their unit cell can be chosen to be inversion symmetric, and therefore one needs at least three different layers in the unit cell to observe such one-dimensional chirality. At the same time, the diagonal components $\eps_{xx}$ and $\eps_{yy}$ of permittivity tensor acquire spatial dispersion corrections which have the second order with respect to ${\bf k}$.

As a result of outlined derivation, the effective permittivity tensor in geometry Fig.~\ref{fig:Structure} takes the following form:
\begin{equation}\label{EffPerm}
\hat{\eps}(\om,{\bf k})=
\begin{pmatrix}
\zeta_0^{-1}+\chi\,k_y^2 & -\varkappa\,k_y+\vartheta\,k_x\,k_y & 0 \\
\varkappa\,k_y+\vartheta\,k_x\,k_y & \eps_{||}-\gamma\,k_y^2 & 0 \\
0 & 0 & \eps_{||}
\end{pmatrix}
\end{equation}
where $\eps_{||}$ is defined by Eq.~\eqref{Epsilon-zz} and the rest of parameters is also defined in terms of Fourier components of permittivity $\eps(x)$ and its inverse $\zeta(x)$:
\begin{gather}
\varkappa=\frac{1}{\zeta_0\,b}\,\sum\limits_{n\not=0}\,\frac{\eps_{-n}\,\zeta_n}{n}\:,\label{Kappa}\\
\chi=\frac{1}{\zeta_0^2\,b^2}\,\sum\limits_{m,n\not=0}\,\frac{\zeta_{-n}\,\eps_{n-m}\,\zeta_{m}}{mn}\:,\label{Chi}\\
\vartheta=-\frac{1}{\zeta_0\,b^2}\,\sum\limits_{n\not=0}\,\frac{\eps_{-n}\zeta_n}{n^2}\:,\label{Theta}\\
\gamma=\zeta_0\,\varkappa^2+\frac{1}{b^2}\,\sum\limits_{m,n\not=0}\,\frac{\eps_{-n}\,\zeta_{n-m}\,\eps_m}{mn}\:.\label{Gamma}
\end{gather}
This particular result Eq.~\eqref{EffPerm} has been obtained in Ref.~\cite{Rizza-2017} for the case of eigenmode propagation in a metamaterial.

Equation~\eqref{EffPerm} is derived for the special case $k_z=0$. We can easily generalize it to the case of arbitrary ${\bf k}$ performing a rotation with respect to $x$ axis and using the transformation law of tensor $\hat{\eps}$ and vector ${\bf k}$: $\hat{\eps}(\om,{\bf k})=\hat{R}\,\hat{\eps}(\om,\hat{R}^{-1}{\bf k})\,\hat{R}^{-1}$, where $\hat{R}$ is the rotation operator. The result reads:
\begin{widetext}
\begin{equation}\label{EffPermFull}
\hat{\eps}(\om,{\bf k})=
\begin{pmatrix}
\zeta_0^{-1}+\chi\,(k_y^2+k_z^2) & -\varkappa\,k_y+\vartheta\,k_x\,k_y & -\varkappa\,k_z+\vartheta\,k_x\,k_z \\
\varkappa\,k_y+\vartheta\,k_x\,k_y & \eps_{||}-\gamma\,k_y^2 & -\gamma\,k_y\,k_z \\
\varkappa\,k_z+\vartheta\,k_x\,k_z & -\gamma\,k_y\,k_z & \eps_{||}-\gamma\,k_z^2
\end{pmatrix}\:.
\end{equation}
\end{widetext}
Tensor~\eqref{EffPermFull} captures the features of metamaterial bulk response including the terms up to the second order in ${\bf k}$ and can be calculated numerically once the profile of permittivity $\eps(x)$ is specified. 

But prior to the analysis of particular situations, we would like to highlight several general properties of the obtained $\hat{\eps}(\om,{\bf k})$, Eq.~\eqref{EffPermFull}.

First, effective permittivity tensor is independent of the unit cell choice. If we shift the coordinate origin within the unit cell by $\Delta$, all Fourier coefficients do change:
\begin{gather}
\tilde{\eps}_n=\eps_n\,e^{inb\,\Delta}\:,\\
\tilde{\zeta}_n=\zeta_n\,e^{inb\,\Delta}\:.
\end{gather}
Hence, all combinations of the form $\eps_n\,\zeta_{-n}$ or $\eps_m\,\zeta_{n-m}\,\eps_{-n}$ or any others with sum of indices equal to zero are essentially independent of the unit cell choice. As such, Eqs.~\eqref{Kappa}--\eqref{Gamma} indicate that the effective permittivity remains unchanged after the shift of the unit cell which is consistent with the general requirement to nonlocal homogenization models~\cite{Alu}.

Second, the effect of one-dimensional chirality described by the $\varkappa$ term vanishes provided the unit cell contains mirror symmetry plane. Without loss of generality we can assume that $x=0$ is a mirror symmetry plane. Then $\eps(x)=\eps(-x)$ and therefore $\eps_n=\eps_{-n}$ and $\zeta_n=\zeta_{-n}$.  As a result, $\eps_n\,\zeta_{-n}=\eps_{-n}\,\zeta_n$ which yields $\varkappa=0$ according to Eq.~\eqref{Kappa}.

Third, in the limit of shallow modulation, i.e. in the situation when $|\eps_n|\ll \eps_0$ for any $n$, all spatial dispersion effects in multilayer are described by the single parameter. To show this, we notice that in this situation $\zeta_0\approx 1/\eps_0$ and $\zeta_n\approx -\eps_n/\eps_0^2$. Using Eqs.~\eqref{Kappa}--\eqref{Gamma}, it is straightforward to check that $\varkappa\approx 0$ and $\chi\approx \vartheta\approx \gamma$.

\section{Assessing local description of multilayers}\label{sec:Permeability}

In many situations it is beneficial to simplify the description of a complex metamaterial to some local model including the set of material parameters which depend only on frequency. An example of such kind is {\it bianisotropic model} which assumes the constitutive relations of the form~\cite{Lindell,Serdyukov}:
\begin{gather}
{\bf D}=\hat{\eps}(\om)\,{\bf E}+\hat{\alpha}(\om)\,{\bf H}\:,\\
{\bf B}=\hat{\beta}(\om)\,{\bf E}+\hat{\mu}(\om)\,{\bf H}\:,
\end{gather}
where $\hat{\eps}$ and $\hat{\mu}$ are local permittivity and permeability tensors, while $\hat{\alpha}$ and $\hat{\beta}$ describe bianisotropic response of the structure. Any medium which fits into the frame of bianisotropic model can be alternatively described by the single nonlocal permittivity tensor~\cite{Skidanov,Silv}:
\begin{equation}\label{NonlocalTensor}
\begin{split}
\hat{\eps}^{\rm{eff}}(\om,{\bf k})=&\left[\hat{\eps}-\hat{\alpha}\hat{\mu}^{-1}\hat{\beta}\right]+\frac{1}{q}\,\left[\hat{\alpha}\hat{\mu}^{-1}{\bf k}^{\times}-{\bf k}^{\times}\hat{\mu}^{-1}\hat{\beta}\right]\\
+&\frac{1}{q^2}\,{\bf k}^{\times}\,\left[\hat{\mu}^{-1}-\hat{I}\right]\,{\bf k}^{\times}\:,
\end{split}
\end{equation}
where ${\bf k}^{\times}$ is an antisymmetric pseudotensor constructed from vector ${\bf k}$ such that ${\bf k}^{\times}\,{\bf a}=[{\bf k}\times {\bf a}]$ for any ${\bf a}$:
\begin{equation}
{\bf k}^{\times}=\begin{pmatrix}
0 & -k_z & k_y \\
k_z & 0 & -k_x \\
-k_y & k_x & 0
\end{pmatrix}
\:.
\end{equation}

Having an explicit expression for the effective permittivity tensor expanded in powers of wave vector, Eq.~\eqref{EffPermFull}, we can analyze whether it can be presented in the form Eq.~\eqref{NonlocalTensor} and, as a consequence, whether local effective material parameters can be introduced. It should be stressed that while spatial dispersion effects may resemble artificial magnetic response for some fixed propagation directions, it does not mean necessarily that local bianisotropic model is adequate for all other propagation directions~\cite{Gorlach-15}.

Analyzing the applicability of local effective medium model, we first note that the structure of the tensors $\hat{\eps}$, $\hat{\mu}$, $\hat{\alpha}$ and $\hat{\beta}$ should be consistent with the symmetry of the metamaterial, i.e. full rotational symmetry with respect to $x$ axis. Therefore, the only possible form of $\hat{\mu}$ is
\begin{equation}\label{PossibleMu}
\hat{\mu}=\begin{pmatrix}
\mu_{\bot} & 0 & 0\\
0 & \mu_{||} & 0 \\
0 & 0 & \mu_{||}
\end{pmatrix}\:,
\end{equation}
i.e. $\hat{\mu}(\omega)$ contains only two independent components. At the same time, second-order spatial dispersion effects in multilayer are described by three independent parameters [see Eq.~\eqref{EffPermFull}]. As we prove in Appendix B, it is not possible to choose such $\mu_{||}$ and $\mu_{\bot}$ that capture all second-order spatial dispersion contributions entering Eq.~\eqref{EffPermFull}. In other words, the bulk properties of multilayered structures cannot be captured by the local bianisotropic model in the general case since there exist second-order spatial dispersion effects beyond such simplified description.

Besides second-order nonlocal effects, multilayered structure also exhibits one-dimensional chirality. We demonstrate below that this particular effect can be described within the frame of the local effective medium model. Assuming $\hat{\mu}=\hat{I}$ in Eq.~\eqref{NonlocalTensor}, we have the following form of the first-order spatial dispersion correction $\delta\eps^{(1)}(\om,{\bf k})$:
\begin{equation}
\delta\eps^{(1)}(\om,{\bf k})=\frac{1}{q}\,\left[\hat{\alpha}\,{\bf k}^{\times}-{\bf k}^{\times}\,\hat{\beta}\right]\:.
\end{equation}
Pseudotensors $\hat{\alpha}$ and $\hat{\beta}$ change their sign under mirror reflection. Therefore, the only possibility to construct such pseudotensor is to use ${\bf n}^{\times}$, where ${\bf n}$ is a unit vector normal to the layers. We assume that $\hat{\alpha}=\hat{\beta}=\varkappa\,q\,{\bf n}^X$, and then
\begin{equation}
\delta\eps^{(1)}=\varkappa\,\begin{pmatrix}
0 & -k_y & -k_z \\
k_y & 0 & 0 \\
k_z & 0 & 0
\end{pmatrix}
\:,
\end{equation} 
which is consistent with the first-order corrections in Eq.~\eqref{EffPermFull}. Thus, bianisotropy of the structure is captured by the following tensors:
\begin{equation}\label{Bianisotropy}
\hat{\alpha}=\hat{\beta}=\begin{pmatrix}
0 & 0 & 0\\
0 & 0 & -\varkappa\,q\\
0 & \varkappa q & 0
\end{pmatrix}
\:,
\end{equation} 
which is so-called omega-type bianisotropy~\cite{Serdyukov}.

\section{Boundary conditions for multilayered structure: layers parallel to the interface}\label{sec:Boundary}

Effective permittivity tensor of multilayered metamaterial, Eq.~\eqref{EffPermFull}, provides full description of wave propagation in an infinite structure. For example, dispersion laws of TE and TM waves for the geometry shown in Fig.~\ref{fig:Structure} read:
\begin{gather}
k_x^2+k_y^2=q^2\,\eps_{||}\mspace{10mu}\:,\label{TEDispesion}\\
\frac{k_x^2}{\eps_{||}}+k_y^2\,\left[\zeta_0-\zeta_0\,q^2\,\chi+\frac{\gamma\,q^2}{\eps_0}-\frac{\varkappa^2\,q^2\,\zeta_0}{\eps_0}\right]\notag\\
+k_x^2\,k_y^2\,\frac{\zeta_0}{\eps_0}\,(\chi+2\,\vartheta)-\frac{\gamma\,\zeta_0\,k_y^4}{\eps_0}\,=q^2\:,\label{TMDispersion}
\end{gather}
respectively. However, to apply the developed theoretical models to any experimental situation, this description has to be supplemented by  appropriate boundary conditions. In this article, we analyze the situation most relevant for the current experiments when the layers are parallel to the interface.

Inspecting dispersion equations Eqs.~\eqref{TEDispesion}, \eqref{TMDispersion}, we find out that each of equations contains the normal component of wave vector  $k_x$ only in the second power. Therefore, $k_x$ is defined uniquely for each of polarizations, and the usual birefringence takes place. As a result, boundary conditions for this situation are obtained from the continuity of tangential components of {\it microscopic} electric and magnetic fields existing in the vicinity of metamaterial boundary aligned with the plane $x=0$. Focusing on TE case, we get: 
\begin{equation}\notag
E_z^{\rm{out}}=E_z(x=0)=E_{0z}+\sum\limits_{n\not=0}\,E_{nz}\:,
\end{equation}
or, making use of calculated Fourier harmonics $E_{nz}$ [Eq.~\eqref{Enz}]
\begin{equation}\label{EteBC}
E_z^{\rm{out}}=\left(1+q^2\,f\right)\,E_{0z}\:,
\end{equation}
where
\begin{equation}\label{FDef}
f=\frac{1}{b^2}\,\sum\limits_{n\not=0}\,\frac{\eps_n}{n^2}\:,
\end{equation}
the superscript ``out'' refers to the fields outside of metamaterial and $E_{0z}$ presents the macroscopic field inside the structure. Equation~\eqref{EteBC} therefore suggests that despite the continuity of microscopic fields, the macroscopic fields do exhibit a discontinuity at the interface. Note that in contrast to the bulk properties the prefactor $f$ [Eq.~\eqref{EteBC}] {\it does depend} on the choice of the boundary and respective unit cell choice.

Next we examine the continuity of microscopic magnetic field
\begin{equation*}
H_y^{\rm{out}}=H_y(x=0)=H_{0y}+\sum\limits_{n\not=0}\,H_{ny}\:,
\end{equation*}
where $H_{ny}=-(k_x+nb)\,E_{nz}/q$ for all $n$. Note that in order to capture second-order spatial dispersion corrections to $H_{ny}$, we need to use the expansion of $E_{nz}$ up to the third order as given by Eq.~\eqref{Enz}. With this expansion, we find out that
\begin{equation}\label{HteBC}
H_y^{\rm{out}}=-\frac{k_x}{q}\,\left(1-q^2\,f\right)\,E_{0z}-q\,g\,E_{0z}\:,
\end{equation}
where 
\begin{equation}\label{GDef}
g=\frac{1}{b}\,\sum\limits_{n\not=0}\,\frac{\eps_n}{n}\:.
\end{equation}
Now in addition to the prefactor $(1-q^2\,f)$ in front of electric field, we also get the term $q\,g$. Both of these terms cause the surface impedance defined for homogenized fields to be different from $-q/k_x$ in the general case. Obviously, this has to be taken into account while retrieving effective parameters of multilayered metamaterial from the transmission and reflection coefficients.

Boundary conditions for TM-polarized waves are derived in a similar way in Appendix C, the result reads:
\begin{gather}
E_y^{\rm{out}}=\frac{H_{0z}}{q}\,\left\lbrace \frac{k_x}{\eps_{||}}+\left(\frac{\varkappa\,\zeta_0}{\eps_0}-\tilde{g}\right)\,k_y^2+\frac{q^2\,f}{\eps_0}\,k_x\notag\right.\\
\left.+\left(\vartheta\,\frac{\zeta_0}{\eps_0}+\frac{\gamma-\zeta_0\,\varkappa^2}{\eps_0^2}-\frac{h}{\eps_0}+\tilde{f}\right)\,k_x\,k_y^2\right\rbrace\:,\label{BoundTM1}\\
H_z^{\rm{out}}=H_{0z}\,\left\lbrace 1+\frac{g\,k_x}{\eps_0}+\left(\frac{g\,\varkappa\,\zeta_0}{\eps_0}-\tilde{h}\right)\,k_y^2-\frac{f}{\eps_0}\,k_x^2\right\rbrace\:,\label{BoundTM2}
\end{gather}
where $f$ and $g$ coefficients are defined by Eqs.~\eqref{FDef}, \eqref{GDef}, while 
\begin{equation}\label{HDef}
h=\frac{1}{b^2}\,\sum\limits_{m,n\not=0}\,\frac{\zeta_{n-m}\,\eps_m}{mn}\:,
\end{equation}
and $\tilde{f}$, $\tilde{g}$ and $\tilde{h}$ are obtained from the respective expressions for $f$, $g$ and $h$ by exchanging $\eps_n$ and $\zeta_n$. Note that in the limit of vanishing spatial dispersion Eqs.~\eqref{BoundTM1}, \eqref{BoundTM2} yield the standard boundary conditions
\begin{gather}
E_y^{\rm{out}}=H_{0z}\,\frac{k_x}{q\,\eps_0}\:,\\
H_z^{\rm{out}}=H_{0z}\:,
\end{gather}
which correspond to the case of a homogeneous medium.

Most remarkably, our results suggest that the first-order spatial dispersion corrections to the boundary conditions exist even in the absence of bulk chirality ($\varkappa=0$) provided the coefficients $g$ and $\tilde{g}$ are nonzero, which is generally the case. Therefore, nonlocal contributions to the boundary conditions may have even more dramatic impact on wave propagation than bulk nonlocality. Note that this first-order spatial dispersion effect was perceived in Ref.~\cite{Popov} as bulk bianisotropy of multilayered structure.

\section{Results for bi-layer structure}\label{sec:Bilayer}

\begin{figure}[t]
\begin{center}
\includegraphics[width=0.95\linewidth]{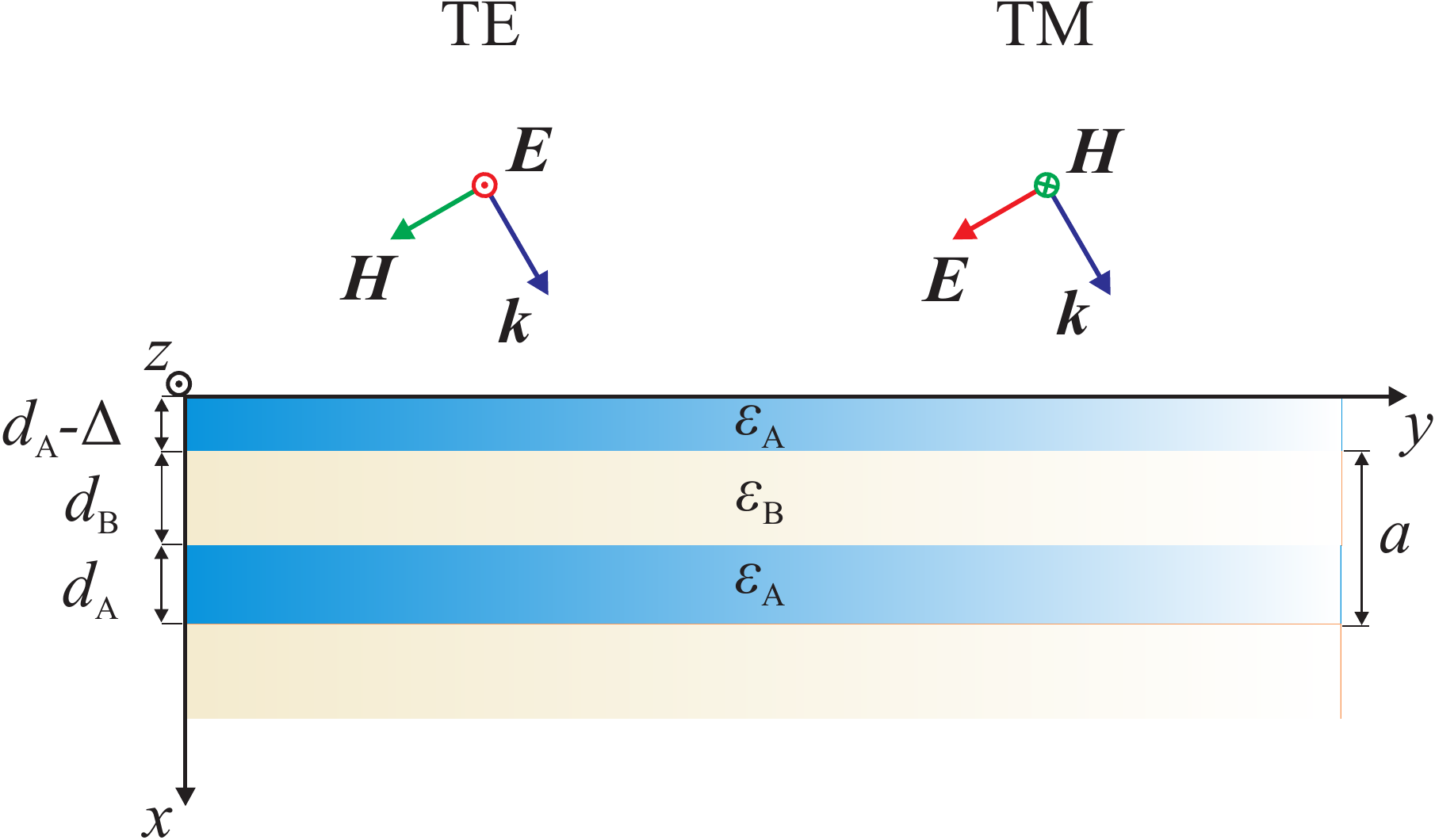}
\caption{The scheme of multilayered structure containing two layers in the unit cell: one with permittivity $\eps_a$ and thickness $d_b$, and another one with permittivity $\eps_b$ and thickness $d_b$. The period of multilayer is equal to $a=d_a+d_b$. In our analysis of boundary problem, we assume that the thickness of the upper layer with permittivity $\eps_a$ is reduced by $\Delta<d_a$. Two possible polarizations of impinging plane wave are sketched, in both cases the wave vector ${\bf k}$ lies in $Oxy$ plane.}
\label{fig:Bilayer}
\end{center}
\end{figure}

To test the developed theoretical models, we consider a simple example of multilayered structure containing two layers in the unit cell with permittivities $\eps_a$ and $\eps_b$ and thicknesses $d_a$ and $d_b$, respectively, with the period of the structure $d_a+d_b=a$ as depicted in Fig.~\ref{fig:Bilayer}. In such case, all parameters which describe spatial dispersion effects in multilayers can be calculated analytically (see Appendix D for details):
\begin{gather}
\eps_{||}=\eps_0+q^2\,(\eps_a-\eps_b)^2\,\frac{d_a^2\,d_b^2}{12\,a^2}\:,\label{EpsTE}\\
\chi=\frac{\eps_0\,(\eps_a-\eps_b)^2}{\zeta_0^2\,\eps_a^2\,\eps_b^2}\,\frac{d_a^2\,d_b^2}{12\,a^2}\:,\label{Chi1}\\
\vartheta=\frac{(\eps_a-\eps_b)^2}{\eps_a\,\eps_b\,\zeta_0}\,\frac{d_a^2\,d_b^2}{12\,a^2}\:,\label{Theta1}\\
\gamma=\zeta_0\,(\eps_a-\eps_b)^2\,\frac{d_a^2\,d_b^2}{12\,a^2}\:,\label{Gamma1}
\end{gather}
where $\eps_0=(\eps_a\,d_a+\eps_b\,d_b)/a$ and $\zeta_0=(\eps_a^{-1}\,d_a+\eps_b^{-1}\,d_b)/a$ are the zeroth order Fourier coefficients of  permittivity $\eps(x)$ and inverse permittivity $\zeta(x)$, respectively. The effect of one-dimensional chirality vanishes in this case, $\varkappa=0$.

The coefficients that enter the boundary conditions read:
\begin{gather}
f=-(\eps_a-\eps_b)\,\frac{d_b}{2\,a}\,\Delta\,\left(\Delta-d_a\right)\notag\\
-(\eps_a-\eps_b)\,\frac{d_a\,d_b}{12\,a}\,(d_a-d_b)\:,\\
g=i\,\left(\eps_a-\eps_b\right)\,\frac{d_b}{a}\,\left(\Delta-\frac{d_a}{2}\right)\:,\\
h=-(\eps_a-\eps_b)\,\frac{d_b}{2\,\eps_a\,a}\,\left(\Delta-\frac{d_a}{2}\right)^2\notag\\
+(\eps_a-\eps_b)\,\frac{d_a\,d_b}{24\,a^2}\,\left[\frac{3\,d_a\,d_b}{\eps_a}+\frac{d_a^2}{\eps_a}+\frac{d_b^2}{\eps_b}\right]\:.
\end{gather}
Similar expressions for $\tilde{f}$, $\tilde{g}$ and $\tilde{h}$ are obtained by replacing $\eps_{a,b}$ by $\eps_{a,b}^{-1}$. 

\begin{figure}[ht!]
\begin{center}
\includegraphics[width=0.85\linewidth]{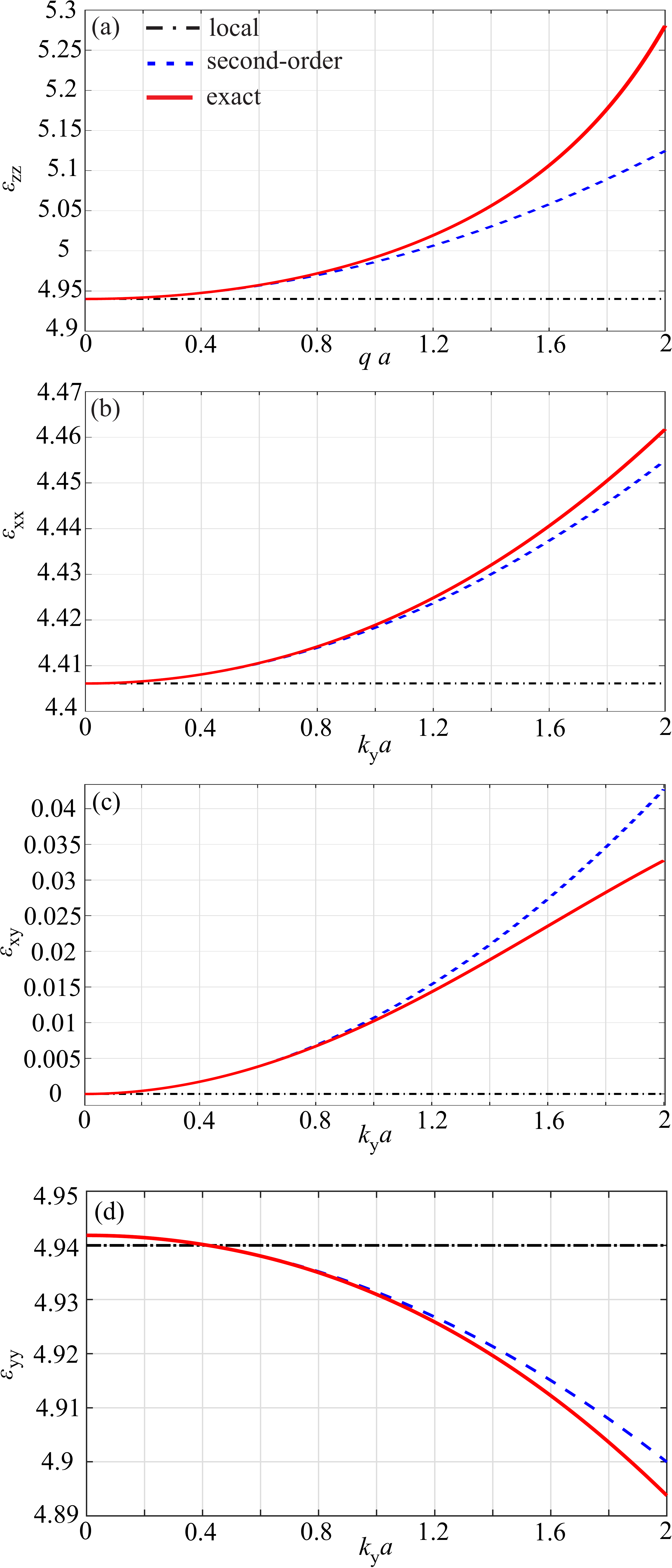}
\caption{Calculated components of the effective nonlocal permittivity tensor for all-dielectric multilayered structure composed of Al$_2$O$_3$ layers with permittivity $\eps_a=3.08$ and TiO$_2$ layers with $\eps_b=6.18$ at wavelengths around $\lambda=1$~$\mu$m. Relative thicknesses of the layers are equal to $d_a=0.4\,a$ and $d_b=0.6\,a$, respectively. Dot-dashed black curves, dashed blue and solid red lines correspond to the predictions of local effective medium model, developed model and exact solution of Ref.~\cite{Chebykin-2011}, respectively.    (a) $\eps_{zz}$ versus dimensionless frequency $q\,a$, where $q=\om/c$. (b-d) Dependence of $\eps_{xx}$, $\eps_{xy}$ and $\eps_{yy}$ on wave number $k_y$ calculated for fixed frequency $qa=0.2$ (wavelength-to-period ratio $\lambda/a\approx 31$) and fixed propagation direction with $k_x=k_y$.}
\label{fig:Permittivity}
\end{center}
\end{figure}

Note that in the special case $\Delta=d_a/2$, i.e. when multilayered structure starts from the layer with half-thickness, the boundary conditions are largely simplified. However, even in such scenario they are nontrivial since $f$ and $\tilde{f}$ coefficients remain nonzero: $f=(\eps_a-\eps_b)\,d_a\,d_b\,(d_a+2\,d_b)/(24\,a)$.

Having the full set of parameters, we first examine the bulk properties of multilayers. To this end, we compare our model Eqs.~\eqref{EffPerm} with parameters Eqs.~\eqref{EpsTE}--\eqref{Gamma1} to the exact though cumbersome  experession for the effective permittivity tensor obtained in Refs.~\cite{Chebykin-2011,Chebykin-12}. As a specific example, we study all-dielectric multilayer composed of Al$_2$O$_3$ and TiO$_2$ layers which has been extensively investigated in recent experiments~\cite{Zhukovsky}.

$\eps_{zz}$ component of permittivity tensor which governs the propagation of TE-polarized waves, exhibits mostly frequency dispersion [Fig.~\ref{fig:Permittivity}(a)], whereas spatial dispersion of $\eps_{zz}$ is quite weak. Predictions of our model nicely match the exact solution up to $q\,a\approx 1$, i.e. $\lambda/a\approx 6.3$, whereas for shorter wavelengths one has to take into account higher-order frequency- and spatial dispersion corrections.

$\eps_{xx}$ and $\eps_{xy}$ components depicted in Fig.~\ref{fig:Permittivity}(b,c) exhibit mostly spatial dispersion and therefore we calculate them for the fixed frequency $qa=0.2$. Again, up to reasonably large wave numbers $k_y\,a\approx 1$ our model provides good accuracy. Note also that the nonzero $\eps_{xy}$ component emerges purely due to spatial dispersion causing small rotation of the multilayer anisotropy axis.

Finally, $\eps_{yy}$ permittivity component [Fig.~\ref{fig:Permittivity}(d)] exhibits both frequency and spatial dispersion, where the former is manifested through the discrepancy between the exact solution and local effective medium model at $k_y\,a\approx 0$, whereas spatial dispersion of $\eps_{yy}$ is well-described by our model up to $k_y\,a\approx 1$.

\begin{figure}[ht!]
\begin{center}
\includegraphics[width=0.78\linewidth]{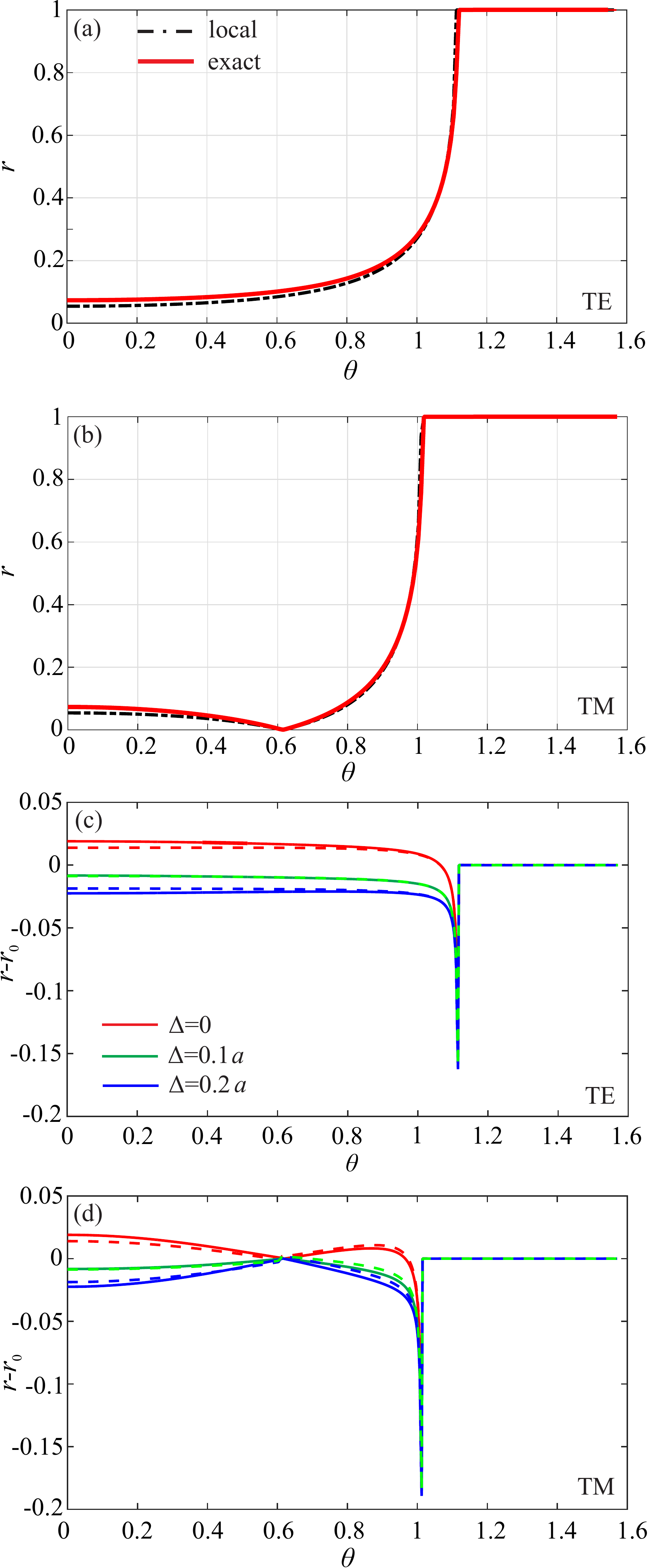}
\caption{Reflectance of a semi-infinite all-dielectric multilayered structure composed of Al$_2$O$_3$ and TiO$_2$ layers with permittivities $\eps_a=3.08$, $\eps_b=6.18$ and thicknesses $d_a=38$~nm ($0.4\,a)$, $d_b=57$~nm ($0.6\,a$), respectively, for the wave incident from ZnSe prism ($\eps_{\rm{out}}=6.14$). Wavelength of incident light $\lambda=1$~$\mu$m, period of the structure $a=95$~nm ($qa=0.6$). (a,b) Reflection coefficient $|E_r/E_{in}|$ ($|H_r/H_{in}|$) for TE (TM)-polarized light. Dot-dashed black and solid red curves show the predictions of local effective medium model and exact results obtained via transfer matrix method, respectively. (c,d) The difference between the reflection coefficients obtained via transfer matrix method (solid lines) or our approach (dashed lines) and reflectance, predicted by the local effective medium model, for TE and TM polarizations. Different colors of the curves correspond to the different metamaterial terminations.}
\label{fig:Reflectance}
\end{center}
\end{figure}

To test the derived boundary conditions, Eqs.~\eqref{EteBC}, \eqref{HteBC}, \eqref{BoundTM1}, \eqref{BoundTM2}, we calculate the reflectance of the same semi-infinite all-dielectric multilayered structure for fixed wavelength and fixed structure parameters as a function of incidence angle $\theta$. To probe the modes of the structure with sufficiently large wave numbers, we consider plane wave incident from ZnSe prism with permittivity $\eps_{\rm{out}}=6.14$ which exceeds both components of multilayer permittivity tensor. So, total internal reflection occurs. Comparison of the local effective medium model with transfer matrix method [Fig.~\ref{fig:Reflectance}(a,b)] yields that the errors of the local effective medium approach are maximal near the angle of total internal reflection, which agrees with experiments~\cite{Zhukovsky}. Nevertheless, as seen from the comparison in Fig.~\ref{fig:Reflectance}(c,d) our model describes the behavior of reflectance quite accurately for all incidence angles.

Apart from the improved accuracy, our model predicts also qualitatively new phenomena not captured by the standard local effective medium model. In Fig.~\ref{fig:Reflectance}(c,d) we compare the reflection coefficients for the same polarization of the incident wave but for the different terminations of multilayered metamaterial. As illustrated in Fig.~\ref{fig:Bilayer}, the difference between these geometries is in the thickness of the upper layer. Note that regardless of the value of $\Delta$ multilayer is strictly periodic with the unit cell containing $d_a-\Delta$ thickness of Al$_2$O$_3$, then $d_b$ layer of TiO$_2$ and finally $\Delta$ layer of TiO$_2$. Therefore, all three configurations with $\Delta=0$, $\Delta=0.1\,a$ and $\Delta=0.2\,a$ essentially correspond to the same metamaterial and differ only by the unit cell shift. As we have proved, bulk properties are unaffected by the unit cell choice and hence the difference in reflectance should be attributed exclusively to the different boundary conditions for these different realizations of the same metamaterial.

It should be stressed that this quite peculiar feature can be potentially tested experimentally providing the direct proof of complicated termination-dependent boundary conditions in multilayers.

\section{Discussion and outlook}\label{sec:Outlook}

In this Article, we have developed a complete framework to describe multilayered structures with arbitrary number of layers in the unit cell, based on the effective medium perspective. Keeping the dominant contributions due to  frequency and spatial dispersion, we have derived both bulk nonlocal effective permittivity tensor, Eq.~\eqref{EffPermFull} and boundary conditions Eqs.~\eqref{EteBC}, \eqref{HteBC} and \eqref{BoundTM1}, \eqref{BoundTM2} for TE- and TM-polarized waves, respectively.

As we have demonstrated, electromagnetic properties of multilayered metamaterials are determined by the complex interplay of the two factors: bulk nonlocality on one side and nonlocal corrections to the boundary conditions (surface nonlocality) on the other. We have proved that the bulk nonlocal response appears to be beyond simplified bianisotropic model based on local permittivity, permeability and bianisotropy tensors. Moreover, surface nonlocality can contain contributions linear with respect to wave vector ${\bf k}$ even in the absence of bulk bianisotropy. This gives rise to rich physics including the dependence of reflectance on termination of multilayered metamaterial. As a consequence, the retrieval of effective material parameters from the measured reflectance and transmittance becomes incorrect unless proper boundary conditions are used. 

To sum up, we believe that our results shed light onto the intricate electromagnetic response of metamaterials suggesting fruitful avenues to design electromagnetic properties desired for applications based on engineering of bulk and surface nonlocalities.

\section*{Acknowledgments}

We acknowledge valuable discussions with Pavel Belov. This work was supported by the Russian Science Foundation (Grant No.~18-72-00102). M.A.G. acknowledges partial support by the Foundation for the Advancement of Theoretical Physics and Mathematics ``Basis''.

\section*{Appendix A. Perturbative analysis of multilayers response}

In this Appendix, we calculate Floquet harmonics of the fields in the multilayered metamaterial applying the perturbation theory with $q\,a$ and $k\,a$ playing the role of small parameters. For clarity, we explicitly introduce small parameter which we denote as $\xi\propto q\,a \propto k\,a$.

{\it TE polarization.~--~}First, we expand the fields as
\begin{gather}
E_{nz}=\xi^2\,E_{nz}^{(2)}+\xi^3\,E_{nz}^{(3)}+\dots\:,\notag\\
D_{nz}=D_{nz}^{(0)}+\xi\,D_{nz}^{(1)}+\dots\:,\notag
\end{gather}
taking into account that the leading term in the expansion of $E_{nz}$ has the second order. Next we rewrite Eqs.~\eqref{FourierTE2} and \eqref{FourierTE4} in the form
\begin{gather}
\left[\left(nb+\xi\,k_x\right)^2+\xi^2\,k_y^2\right]\,\left[\xi^2\,E_{nz}^{(2)}+\xi^3\,E_{nz}^{(3)}+\dots\right]\notag\\
=\xi^2\,q^2\,\left(D_{nz}^{(0)}+\xi\,D_{nz}^{(1)}+\dots\right)\:,\\
D_{nz}^{(0)}+\xi\,D_{nz}^{(1)}+\dots=\eps_n\,E_{0z}\notag\\
+\sum\limits_{m\not=-n}\,\eps_{-m}\,\left(\xi^2\,E_{(m+n)z}^{(2)}+\xi^3\,E_{(m+n)z}^{(3)}+\dots\right)\:.
\end{gather}
Separating the equations for different orders of $\xi$, we recover that
\begin{gather}
n^2\,b^2\,E_{nz}^{(2)}=q^2\,D_{nz}^{(0)}\:,\\
2\,nb\,k_x\,E_{nz}^{(2)}+n^2\,b^2\,E_{nz}^{(3)}=q^2\,D_{nz}^{(1)}\:,\\
D_{nz}^{(0)}=\eps_n\,E_{0z}\:,\\
D_{nz}^{(1)}=0\:,
\end{gather}
which eventually yield Eq.~\eqref{Enz} of the article main text:
\begin{equation*}
E_{nz}=\frac{q^2\,\eps_n}{n^2\,b^2}\,E_{0z}-\frac{2k_x\,q^2\,\eps_n}{n^3\,b^3}\,E_{0z}\:.
\end{equation*}
Note that this third-order expansion is necessary for the correct derivation of the boundary condition for magnetic field, Eq.~\eqref{HteBC}.

{\it TM polarization.~--}~Using Eq.~\eqref{WaveEq} together with the material equation ${\bf D}({\bf r})=\eps(x)\,{\bf E}({\bf r})$, we get:
\begin{gather}
k_y^2\,E_{nx}-k_x^{(n)}\,k_y\,E_{ny}=q^2\,D_{nx}\mspace{10mu} (n\not=0)\:,\label{FourierTM1}\\
-k_x^{(n)}\,k_y\,E_{nx}+\left[k_x^{(n)}\right]^2\,E_{ny}=q^2\,D_{ny}\mspace{10mu} (n\not=0)\:,\label{FourierTM2}\\
E_{nx}=\zeta_n\,D_{0x}+\sum\limits_{m\not=0}\,\zeta_{n-m}\,D_{mx}\:,\label{FourierTM3}\\
D_{ny}=\eps_n\,E_{0y}+\sum\limits_{m\not=0}\,\eps_{n-m}\,E_{my}\:.\label{FourierTM4}
\end{gather}
In this system, we omit two equations for $E_{0x}$ and $E_{0y}$ Floquet harmonics which are also the consequence of the wave equation and yield the relation between $E_{0x}$ and $E_{0y}$ from one side and current densities $J_x$ and $J_y$ from the other. These equations are not especially useful in $\hat{\eps}(\om,{\bf k})$ calculation, though they are needed in the calculation of the Green's function of multilayered structure. Note also that since $\divv\,{\bf D}=0$,
\begin{equation}\label{Gauss}
k_x^{(n)}\,D_{nx}+k_y\,D_{ny}=0\:,
\end{equation}
which can be derived by combining Eqs.~\eqref{FourierTM1} and \eqref{FourierTM2}.

In a fully static case $q=k=0$ we get $E_{ny}=0$ ($n\not=0$) [Eq.~\eqref{FourierTM2}] and also $D_{nx}=0$ ($n\not=0$) [Eq.~\eqref{Gauss}]. As a consequence of that, Eq.~\eqref{FourierTM3} yields $E_{0x}=\zeta_0\,D_{0x}$, $D_{0y}=\eps_0\,E_{0x}$. Hence, $\eps^{\rm{loc}}_{xx}=\zeta_0^{-1}$ and $\eps^{\rm{loc}}_{yy}=\eps_0$ in the quasistatic limit.

Now we seek second-order spatial dispersion corrections to these formulas treating $E_{ny}$ and $D_{nx}$ as small parameters. We rewrite the Eqs.~\eqref{FourierTM1}--\eqref{FourierTM4} as follows:
\begin{gather}
E_{nx}=\zeta_n\,D_{0x}+\sum\limits_{m\not=0}\,\zeta_{n-m}\,D_{mx}\:,\label{TM1}\\
D_{ny}=\eps_n\,E_{0y}+\sum\limits_{m\not=0}\,\eps_{n-m}\,E_{my}\:,\label{TM2}\\
E_{ny}=\frac{\xi\,k_y}{nb+\xi\,k_x}\,E_{nx}+\frac{\xi^2\,q^2}{\left[nb+\xi\,k_x\right]^2}\,D_{ny} \mspace{10mu}(n\not=0)\:,\label{TM3}\\
D_{nx}=-\frac{\xi\,k_y}{nb+\xi\,k_x}\,D_{ny}\mspace{10mu} (n\not=0)\:.\label{TM4}
\end{gather}
Next we expand all Floquet harmonics with $n\not=0$ in terms of the small parameter $\xi$:
\begin{gather}
E_{nx}=E_{nx}^{(0)}+\xi\,E_{nx}^{(1)}+\xi^2\,E_{nx}^{(2)}+\dots\:,\label{Exp1}\\
D_{ny}=D_{ny}^{(0)}+\xi\,D_{ny}^{(1)}+\xi^2\,D_{ny}^{(2)}+\dots\:,\label{Exp2}\\
E_{ny}=\xi\,E_{ny}^{(1)}+\xi^2\,E_{ny}^{(2)}+\dots\:,\label{Exp3}\\
D_{nx}=\xi\,D_{nx}^{(1)}+\xi^2\,D_{nx}^{(2)}+\dots\:\label{Exp4}
\end{gather}
and calculate the expansion coefficients iteratively. Based on Eqs.~\eqref{TM1}, \eqref{TM2},
\begin{gather}
E_{nx}^{(0)}=\zeta_n\,D_{0x}\:,\\
D_{ny}^{(0)}=\eps_n\,E_{0y}\:.
\end{gather}
Using Eqs.~\eqref{TM3} and \eqref{TM4} we derive:
\begin{gather}
E_{ny}^{(1)}=\frac{k_y}{nb}\,E_{nx}^{(0)}=\frac{k_y\,\zeta_n}{nb}\,D_{0x}\:,\label{Eny1}\\
D_{nx}^{(1)}=-\frac{k_y}{nb}\,D_{ny}^{(0)}=-\frac{k_y\,\eps_n}{nb}\,E_{0y}\:.\label{Dnx1}
\end{gather}
Returning with these results to Eqs.~\eqref{TM1}, \eqref{TM2}, we get:
\begin{gather}
E_{nx}^{(1)}=\sum\limits_{m\not=0}\,\zeta_{n-m}\,D_{mx}^{(1)}=-\sum\limits_{m\not=0}\,\zeta_{n-m}\,\frac{k_y\,\eps_m}{mb}\,E_{0y}\:,\\
D_{ny}^{(1)}=\sum\limits_{m\not=0}\,\eps_{n-m}\,E_{my}^{(1)}=\sum\limits_{m\not=0}\,\eps_{n-m}\,\frac{k_y\,\zeta_m}{mb}\,D_{0x}\:.
\end{gather}
Applying Eqs.~\eqref{TM3}, \eqref{TM4} once again, we recover that
\begin{gather}
E_{ny}^{(2)}=-\frac{k_y\,k_x}{n^2\,b^2}\,E_{nx}^{(0)}+\frac{q^2}{n^2\,b^2}\,D_{ny}^{(0)}+\frac{k_y}{nb}\,E_{nx}^{(1)}\notag\\
=-\frac{k_y\,k_x}{n^2\,b^2}\,\zeta_n\,D_{0x}+\left[\frac{q^2\eps_n}{n^2b^2}-\frac{k_y^2}{nb}\,\sum\limits_{m\not=0}\frac{\zeta_{n-m}\eps_m}{mb}\right]\,E_{0y}\:,\label{Eny2}\\
D_{nx}^{(2)}=-\frac{k_y}{nb}\,D_{ny}^{(1)}+\frac{k_y\,k_x}{n^2\,b^2}\,D_{ny}^{(0)}\notag\\
=-\frac{k_y^2}{b^2}\,\sum\limits_{m\not=0}\,\frac{\eps_{n-m}\,\zeta_m}{nm}\,D_{0x}+\frac{k_x\,k_y}{n^2\,b^2}\,\eps_n\,E_{0y}\:.\label{Dnx2}
\end{gather}

Then we use Eqs.~\eqref{TM1}, \eqref{TM2}, but with $n=0$:
\begin{gather}
E_{0x}=\zeta_0\,D_{0x}+\sum\limits_{n\not=0}\,\zeta_{-n}\,\left[D_{nx}^{(1)}+D_{nx}^{(2)}\right]\:,\label{TMFinal1}\\
D_{0y}=\eps_0\,E_{0y}+\sum\limits_{n\not=0}\,\eps_{-n}\,\left[E_{ny}^{(1)}+E_{ny}^{(2)}\right]\:,\label{TMFinal2}
\end{gather}
In the right-hand side of these equations, we use the expressions \eqref{Eny1}, \eqref{Dnx1}, \eqref{Eny2}, \eqref{Dnx2} and deduce:
\begin{gather}
E_{0x}=\left(\zeta_0-\zeta_0^2\,\chi\,k_y^2\right)\,D_{0x}\notag\\
+\left(\zeta_0\,\varkappa\,k_y-\zeta_0\,\vartheta\,k_x\,k_y\right)\,E_{0y}\:,\label{E0x}\\
D_{0y}=\left[\eps_{||}+(\zeta_0\,\varkappa^2-\gamma)\,k_y^2\right]\,E_{0y}\notag\\
+\left[\zeta_0\,\varkappa\,k_y+\zeta_0\,\vartheta\,k_x\,k_y\right]\,D_{0x}\:,\label{D0y}
\end{gather}
where we have used the designations Eqs.~\eqref{Kappa}-\eqref{Gamma}. Solving Eqs.~\eqref{E0x},\eqref{D0y} with respect to $D_{0x}$ and $D_{0y}$, we arrive to the following set of constitutive relations:
\begin{gather}
D_{0x}=\left(\zeta_0^{-1}+\chi\,k_y^2\right)\,E_{0x}+(-\varkappa\,k_y+\vartheta\,k_x\,k_y)\,E_{0y}\:,\label{D0xfin}\\
D_{0y}=\left(\varkappa\,k_y+\vartheta\,k_x\,k_y\right)\,E_{0x}+(\eps_{||}-\gamma\,k_y^2)\,E_{0y}\:.\label{D0yfin}
\end{gather}
In this way, effective permittivity tensor in the chosen geometry reads:
\begin{equation*}
\hat{\eps}(\om,{\bf k})=\begin{pmatrix}
\zeta_0^{-1}+\chi\,k_y^2 & -\varkappa\,k_y+\vartheta\,k_x\,k_y & 0 \\
\varkappa\,k_y+\vartheta\,k_x\,k_y & \eps_{||}-\gamma\,k_y^2 & 0\\
0 & 0 & \eps_{||}
\end{pmatrix}
\:,
\end{equation*}
which is Eq.~\eqref{EffPerm} provided in the article main text.

\section*{Appendix B. On applicability of local permeability tensor to describe multilayered structures}

Multilayered structure which we study has full rotational symmetry with respect to $x$ axis. Therefore, the symmetry restricts possible form of permeability tensor and guarantees that it can only include identity matrix $\hat{I}$ and the dyadics ${\bf n}\otimes{\bf n}$, where ${\bf n}$ is a unit vector normal to the layers:
\begin{equation}
\frac{1}{q^2}\,\left[\hat{\mu}^{-1}-\hat{I}\right]\equiv \hat{A}=\begin{pmatrix}
A_1 & 0 & 0\\
0 & A_2 & 0 \\
0 & 0 & A_2
\end{pmatrix}\:.
\end{equation}
In such a case, local effective medium model would demand the following form of the second-order spatial dispersion correction $\delta\eps^{(2)}$:
\begin{equation}\label{Magnetic}
\begin{split}
& \delta\eps^{(2)}(\om,{\bf k})={\bf k}^{\times}\,\hat{A}\,{\bf k}^{\times}\\
& =
\begin{pmatrix}
-A_2\,(k_y^2+k_z^2) & A_2\,k_x\,k_y & A_2\,k_x\,k_z \\
A_2\,k_x\,k_y & -A_2\,k_x^2-A_1\,k_z^2 & A_1\,k_y\,k_z \\
A_2\,k_x\,k_z & A_1\,k_y\,k_z & -A_2\,k_x^2-A_1\,k_y^2
\end{pmatrix}
\:.
\end{split}
\end{equation} 
However, as we have shown, second-order spatial dispersion effects in multilayers are described by
\begin{equation}\label{DeltaEps2}
\delta\eps^{(2)}(\om,{\bf k})=
\begin{pmatrix}
\chi\,(k_y^2+k_z^2) & \vartheta\,k_x\,k_y & \vartheta\,k_x\,k_z \\
\vartheta\,k_x\,k_y & -\gamma\,k_y^2 & -\gamma\,k_y\,k_z \\
\vartheta\,k_x\,k_z & -\gamma\,k_y\,k_z & -\gamma\,k_z^2
\end{pmatrix}
\:.
\end{equation} 
Equations~\eqref{Magnetic} and \eqref{DeltaEps2} appear to be incompatible, though we may ensure the same off-diagonal entries by choosing $A_1=-\gamma$ and $A_2=\vartheta$. However, the diagonal entries will be different, even in the shallow modulation limit, since they include different components of wave vector.

Therefore, second-order spatial dispersion effects in multilayers generally can not be described in terms of local permeability tensor. This feature is manifested in a number of physical phenomena, e.g. tri-refringence~\cite{Orlov}.

\section*{Appendix C. Derivation of boundary conditions for TM-polarized waves}

Here, we derive boundary conditions for TM-polarized waves incident on the surface of a multilayered metamaterial with layers parallel to the interface.

First, Maxwell's equations yield
\begin{equation}\label{D0xy}
D_{0x}=-\frac{k_y}{q}\,H_{0z}\:,\mspace{8mu} D_{x}^{\rm{out}}=-\frac{k_y}{q}\,H_z^{\rm{out}}\:,\mspace{8mu} D_{0y}=\frac{k_x}{q}\,H_{0z}\:.
\end{equation}
Second, combining Eqs.~\eqref{D0y} and \eqref{D0xy}, we recover that
\begin{equation}\label{E0y-H0z}
E_{0y}=\frac{H_{0z}}{q}\,\left\lbrace \frac{k_x}{\eps_{||}}+\frac{\varkappa\,\zeta_0}{\eps_0}\,k_y^2+\frac{\vartheta\,\zeta_0\,\eps_0+\gamma-\zeta_0\,\varkappa^2}{\eps_0^2}\,k_x\,k_y^2\right\rbrace\:.
\end{equation}
Now, we use the continuity of miroscopic fields at metamaterial boundary:
\begin{gather}
E_y^{\rm{out}}=E_{0y}+\sum\limits_{n\not=0}\,E_{ny}\:,\label{TMCont1}\\
D_x^{\rm{out}}=D_{0x}+\sum\limits_{n\not=0}\,D_{nx}\:,\label{TMCont2}
\end{gather}
where $E_{ny}$ and $D_{nx}$ are defined by Eqs.~\eqref{Eny1}, \eqref{Eny2} and \eqref{Dnx1}, \eqref{Dnx2}, respectively. As a result, boundary conditions take the form
\begin{gather}
E_y^{\rm{out}}=E_{0y}\,\left(1+q^2\,f-k_y^2\,h\right)+D_{0x}\,\left(k_y\,\tilde{g}-k_x\,k_y\,\tilde{f}\right)\:,\label{BoundPrelim1}\\
D_x^{\rm{out}}=E_{0y}\,\left(-k_y\,g+k_x\,k_y\,f\right)+D_{0x}\,\left(1-k_y^2\,\tilde{h}\right)\label{BoundPrelim2}
\end{gather}
with $f$, $g$, $h$ coefficients defined by Eqs.~\eqref{FDef}, \eqref{GDef} and \eqref{HDef}, respectively. The expressions for $\tilde{f}$, $\tilde{g}$ and $\tilde{h}$ are obtained by replacing $\eps_n$ by $\zeta_n$ and vice versa. Making use of Eqs.~\eqref{D0xy}, \eqref{E0y-H0z}, we convert Eqs.~\eqref{BoundPrelim1}-\eqref{BoundPrelim2} to their final form Eqs.~\eqref{BoundTM1}-\eqref{BoundTM2}.

\section*{Appendix D. Calculations for bi-layer structure}

In this Appendix, we outline the technique to calculate the parameters characterizing bulk and surface properties of multilayered structure, doing this calculation explicitly for bi-layer structure. Our approach is based on the conversion of sums Eqs.~\eqref{Kappa}--\eqref{Gamma}, \eqref{FDef}, \eqref{GDef} and \eqref{HDef} involving Fourier coefficients of permittivity and its inverse into the real-space integrals which are much easier to calculate. To this end, we introduce a set of functions with zero average:
\begin{gather}
\eps^{(0)}(x)\equiv \eps(x)-\eps_0=\sum\limits_{n\not=0}\,\eps_n\,e^{inbx}\:,\\
\eps^{(1)}(x)=\int\,\eps^{(0)}(x)\,dx=\sum\limits_{n\not=0}\,\frac{\eps_n}{inb}\,e^{inbx}\:,\\
\eps^{(2)}(x)=\int\,\eps^{(1)}(x)\,dx=\sum\limits_{n\not=0}\,-\frac{\eps_n}{n^2}\,e^{inbx}\:.
\end{gather}
In a similar way we define $\zeta^{(0)}(x)$, $\zeta^{(1)}(x)$ and $\zeta^{(2)}(x)$. Obviously,
\begin{gather}
\eps^{(0)}(x)=
\begin{cases}
(\eps_a-\eps_b)\,d_b/a, & 0<x<d_a\:,\\
-(\eps_a-\eps_b)\,d_a/a, & d_a<x<d_b\:.
\end{cases}
\end{gather}
Integrating these expressions, we derive that
\begin{gather}
\eps^{(1)}(x)=
\begin{cases}
(\eps_a-\eps_b)\,\frac{d_b}{a}\,\left(x-\frac{d_a}{2}\right), & 0<x\leq d_a\:,\\
(\eps_a-\eps_b)\,\frac{d_a}{a}\,\left(a-x-\frac{d_b}{2}\right), & d_a\leq x<d_b\:,
\end{cases}
\\
\eps^{(2)}(x)=\frac{(\eps_a-\eps_b)\,d_b}{2\,a}\,\left(x-\frac{d_a}{2}\right)^2\notag\\
-\frac{(\eps_a-\eps_b)\,d_a\,d_b}{24\,a}\,(a+d_b)\: \text{ for }\: 0<x\leq d_a\:,\\
\eps^{(2)}(x)=-\frac{(\eps_a-\eps_b)\,d_a}{2\,a}\,\left(a-x-\frac{d_b}{2}\right)^2\notag\\
+\frac{(\eps_a-\eps_b)\,d_a\,d_b}{24\,a}\,(a+d_a)\: \text{ for }\: d_a\leq x<d_b\:.
\end{gather}
The functions $\zeta^{(0)}(x)$, $\zeta^{(1)}(x)$ and $\zeta^{(2)}(x)$ are obtained from these expressions by replacing $\eps_{a,b}$ by $\eps_{a,b}^{-1}$. Parameters of multilayered structure are defined in terms of the introduced functions as follows:
\begin{gather}
\eps_{||}=\eps_0+\frac{q^2}{a}\,\int\limits_0^a\,\left[\eps^{(1)}(x)\right]^2\,dx\:,\\
\varkappa=\frac{i}{\zeta_0\,a}\,\int\limits_0^a\,\eps^{(0)}(x)\,\zeta^{(1)}(x)\,dx\:,\\
\chi=\frac{1}{\zeta_0^2\,a}\,\int\limits_0^a\,\eps(x)\,\left[\zeta^{(1)}(x)\right]^2\,dx\:,\\
\vartheta=-\frac{1}{\zeta_0\,a}\,\int\limits_0^a\,\eps^{(1)}(x)\,\zeta^{(1)}(x)\,dx\:,\\
\gamma=\zeta_0\,\varkappa^2+\frac{1}{a}\,\int\limits_0^a\,\left[\eps^{(1)}(x)\right]^2\,\zeta(x)\,dx\:,\\
f=-\eps^{(2)}(\Delta)\:,\label{Fdef2}\\
g=i\,\eps^{(1)}(\Delta)\:,\label{Gdef2}\\
h=-\mathfrak{H}^{(1)}(\Delta)\:,\label{Hdef2}
\end{gather}
and $\tilde{f}$, $\tilde{g}$ and $\tilde{h}$ coefficients are obtained by replacing the permittivity in Eqs.~\eqref{Fdef2}-\eqref{Hdef2} by its inverse. Here, the functions $\mathfrak{H}^{(0)}(x)$ and $\mathfrak{H}^{(1)}(x)$ are defined as:

\begin{gather}
\mathfrak{H}^{(0)}(x)\equiv\zeta(x)\,\eps^{(1)}(x)-\left\langle \zeta(x)\,\eps^{(1)}(x)\right\rangle\notag\\
=\begin{cases}
(\eps_a-\eps_b)\,\frac{d_b}{\eps_a\,a}\,\left(x-\frac{d_a}{2}\right), & 0<x<d_a\:,\\
(\eps_a-\eps_b)\,\frac{d_a}{\eps_b\,a}\,\left(a-x-\frac{d_b}{2}\right), & d_a<x<d_b\:,
\end{cases}
\label{H0def}
\end{gather}
where $\left\langle \dots \right\rangle$ denotes the average over the unit cell.

\begin{equation*}
\mathfrak{H}^{(1)}(x)=\int\,\mathfrak{H}^{(0)}(x)\,dx+C\:,
\end{equation*}
with  the integration constant chosen in such way that the average of $\mathfrak{H}^{(1)}(x)$ is equal to zero.

\section*{Appendix E. Calculation of reflectance}

Calculating the reflectance of a semi-infinite multilayered structure within our model, we perform the following conceptual steps. We define wave vector components for the incident wave: $k_x^{\rm{in}}=q\,\sqrt{\eps_{\rm{out}}}\,\cos\theta$, $k_y=q\,\sqrt{\eps_{\rm{out}}}\,\sin\theta$. 

{\it TE-polarized waves.~--~} $k_x$ for the transmitted wave is found from the dispersion equation Eq.~\eqref{TEDispesion}:
\begin{equation}
k_x=\sqrt{q^2\,\eps_{||}-k_y^2}\:.
\end{equation}
Also we take into account the link between electric and magnetic fields for the incident as well as for the reflected waves:
\begin{equation}
H_y^{\rm{in,r}}=\mp\frac{k_x^{\rm{in}}}{q}\,E_z^{\rm{in,r}}\:.
\end{equation}
Then, applying the boundary conditions Eqs.~\eqref{EteBC}, \eqref{HteBC}, we finally derive:
\begin{equation}\label{ReflTE}
r^{\rm{TE}}\equiv \frac{E_z^{\rm{r}}}{E_z^{\rm{in}}}=\frac{k_x^{\rm{in}}\,(1+q^2\,f)-k_x\,(1-q^2\,f)-q^2\,g}{k_x^{\rm{in}}\,(1+q^2\,f)+k_x\,(1-q^2\,f)+q^2\,g}\:.
\end{equation}

{\it TM-polarized waves.~--~} In a similar way we analyze the boundary problem for TM-polarized waves. $k_x$ for the transmitted wave is defined uniquely by Eq.~\eqref{TMDispersion}. The link between electric and magnetic fields for the incident as well as for the reflected waves reads:
\begin{equation}
E_y^{\rm{in,r}}=\pm\frac{k_x^{\rm{in}}}{q\,\eps_{\rm{out}}}\,H_z^{\rm{in,r}}\:.
\end{equation}
Applying the boundary conditions Eqs.~\eqref{BoundTM1}, \eqref{BoundTM2}, we find the reflection coefficient in the form
\begin{equation}
r^{\rm{TM}}\equiv\frac{H_z^{\rm{r}}}{H_z^{\rm{in}}}=\frac{k_x^{\rm{in}}\,B-\eps_{\rm{out}}\,A}{k_x^{\rm{in}}\,B+\eps_{\rm{out}}\,A}\:,
\end{equation}
where $A$ and $B$ coefficients come from the boundary conditions and read:
\begin{gather}
A=\frac{k_x}{\eps_{||}}+\left(\frac{\varkappa\,\zeta_0}{\eps_0}-\tilde{g}\right)\,k_y^2+\frac{q^2\,f}{\eps_0}\,k_x\notag\\
+\left(\vartheta\,\frac{\zeta_0}{\eps_0}+\frac{\gamma-\zeta_0\,\varkappa^2}{\eps_0^2}-\frac{h}{\eps_0}+\tilde{f}\right)\,k_x\,k_y^2\:,\\
B=1+\frac{g\,k_x}{\eps_0}+\left(\frac{g\,\varkappa\,\zeta_0}{\eps_0}-\tilde{h}\right)\,k_y^2-\frac{f}{\eps_0}\,k_x^2\:.
\end{gather}
The dependence of reflectance on multilayer termination arises due to the dependence of $f$, $g$, $h$, $\tilde{f}$, $\tilde{g}$ and $\tilde{h}$ coefficients on the thickness of the upper layer, $d_a-\Delta$.

{\it Transfer matrix method.~--} Transfer matrix method for wave propagation in stratified media is analyzed in detail in the classical textbook~\cite{Born}, here we outline the main calculation steps. Transfer matrices defined for TE and TM waves as
\begin{gather}
\begin{pmatrix}
E_z(x)\\
H_y(x)
\end{pmatrix}
=M_{\rm{TE}}\,
\begin{pmatrix}
E_z(0)\\
H_y(0)
\end{pmatrix}
\\
\begin{pmatrix}
E_y(x)\\
H_z(x)
\end{pmatrix}
=M_{\rm{TM}}\,
\begin{pmatrix}
E_y(0)\\
H_z(0)
\end{pmatrix}\:,
\end{gather}
in the medium with the diagonal permittivity tensor have the form
\begin{gather}
M_{\rm{TE}}(x)=\begin{pmatrix}
\cos k_x\,x & -\frac{iq}{k_x}\,\sin k_x x\\
-\frac{i\,k_x}{q}\,\sin k_x x & \cos k_x x
\end{pmatrix}
\:,\\
M_{\rm{TM}}(x)=\begin{pmatrix}
\cos k_x\,x & \frac{ik_x}{q\,\eps_{yy}}\,\sin k_x\,x\\
\frac{iq\,\eps_{yy}}{k_x}\,\sin k_x x & \cos k_x\,x
\end{pmatrix}
\:,
\end{gather}
where $x$ axis is chosen as the propagation direction. Constructing $M=M(\eps_a,\Delta)\,M(\eps_b,d_b)\,M(\eps_a,d_a-\Delta)$, we obtain the transfer matrix for a single period of a metamaterial. The transmitted wave satisfies the equation
\begin{equation}\label{BlochMode}
\begin{pmatrix}
m_{11}-e^{ik_x\,x} & m_{12}\\
m_{21} & m_{22}-e^{ik_x\,x}
\end{pmatrix}
\,
\begin{pmatrix}
E^{\rm{t}}\\
H^{\rm{t}}
\end{pmatrix}
=0
\:,
\end{equation}
where $k_x$ is a Bloch wave number of the transmitted wave, $E^{\rm{t}}$ and $H^{\rm{t}}$ are the tangential components of the transmitted Bloch wave directly near the boundary of a metamaterial, and $m_{ij}$ are the elements of the constructed transfer matrix $M$. From this equation, we immediately evaluate the impedance and admittance of the Bloch wave:
\begin{equation}\label{BlochWaveImp}
Z_{\rm{t}}\equiv \frac{E^{\rm{t}}}{H^{\rm{t}}}=-\frac{m_{22}-e^{ik_x\,x}}{m_{21}}\:,
\end{equation}
$Y_{\rm{t}}\equiv Z_{\rm{t}}^{-1}$. The reflection coefficients from the semi-infinite medium are then found as
\begin{gather}
r_{\rm{TE}}\equiv\frac{E_z^{\rm{r}}}{E_z^{\rm{in}}}=\frac{Y_{\rm{in}}^{\rm{TE}}-Y_{\rm{t}}^{\rm{TE}}}{Y_{\rm{in}}^{\rm{TE}}+Y_{\rm{t}}^{\rm{TE}}}\:,\\
r_{\rm{TM}}\equiv\frac{H_z^{\rm{r}}}{H_z^{\rm{in}}}=\frac{Z_{\rm{in}}^{\rm{TM}}-Z_{\rm{t}}^{\rm{TM}}}{Z_{\rm{in}}^{\rm{TM}}+Z_{\rm{t}}^{\rm{TM}}}\:,
\end{gather}
where $Y_{\rm{in}}^{\rm{TE}}=-k_x^{\rm{in}}/q$ and $Z_{\rm{in}}^{\rm{TM}}=k_x^{\rm{in}}/(\eps_{\rm{out}}\,q)$.

%
%
%
%
%
%
%
%

\bibliography{NLMultilayer}

\end{document}